\def \myplotone#1 {\centerline{#1}}
\def \myplotfiddle#1 {\centerline{#1}}

\def \etal {{\it et al. }}

\def \farcs{\hbox{$.\!\!^{\prime\prime}$}}
\def \farcm{\hbox{$.\!\!^{\prime}$}}

\documentstyle[aaspp4,11pt]{article}
\eqsecnum
\slugcomment{Submitted to Ap.J.}

\begin{document}

\title{The Dark Matter, Gas and Galaxy Distributions in Abell~2218:\\
 A Weak Gravitational Lensing and X-ray Analysis}
\author{G. Squires\altaffilmark{1}
N. Kaiser\altaffilmark{2}\altaffilmark{,6},
A. Babul\altaffilmark{3},
G. Fahlman\altaffilmark{4}\altaffilmark{,6},
D. Woods\altaffilmark{4},
D. M. Neumann\altaffilmark{5} and H. B\"ohringer\altaffilmark{5}}
\altaffiltext{1}{Physics Department, University of Toronto\\
60 St.\ George St., Toronto, Ontario, Canada M5S 1A7}
\altaffiltext{2}{Canadian Institute for Advanced Research and\\
Canadian Institute for Theoretical Astrophysics, University of Toronto\\
60 St.\ George St., Toronto, Ontario, Canada M5S 1A7}
\altaffiltext{3}{Department of Physics, New York University\\
4 Washington Place, Room 525, New York, New York, USA 1003-6621}
\altaffiltext{4}{Department of Geophysics and Astronomy, University of
British Columbia\\ 2219, Main Mall, Vancouver, BC, Canada V6T 1Z4}
\altaffiltext{5}{Max-Planck-Institut f\"ur extraterrestrische Physik\\
D-85740 Garching bei M\"unchen, Federal Republic of Germany}
\altaffiltext{6}{Visiting Astronomer, Canada-France Hawai'i Telescope.
Operated by the:\\ National Research Council of Canada, le Centre National
de la Recherche\\ Scientifique de France and the University of Hawai'i}

\begin{abstract}
We report on the detection of dark matter in the cluster Abell 2218 using
the weak gravitational distortion of background galaxies. We find
a highly significant, coherent  detection of the distortion in the
images of the background galaxies. We use HST images from the Medium
Deep Survey to calibrate the suppression in the observed distortion due to
atmospheric smearing. The inferred 2D mass distribution has a peak that
is coincident with the optical and X-ray centroid. The qualitative
distributions of the cluster light, the X-ray emission and the dark matter are
similar and the projected total mass, gas, and light surface densities
are  consistent with a $r^{-1}$ profile at distance of
$r > 180^{\prime \prime}$ from the cluster cD galaxy.
Using the weak lensing technique, we determine a lower bound for the
total mass in A2218 of $(3.9 \pm 0.7) \times 10^{14}$~h$^{-1}$~M$_\odot$ within
a fiducial aperture of radius 0.4~h$^{-1}$Mpc.
The associated cluster
mass-to-light ratio is $(440 \pm 80)$~h~$M_\odot/L_{\odot B}$.
The mass estimated by the weak lensing method is consistent with that
inferred from the X-ray data under the assumption of hydrostatic equilibrium
and we derive an upper bound
for the gas-to-total mass ratio at 400~h$^{-1}$kpc of $M_{gas}/M_{tot} =
(0.04 \pm 0.02)$h$^{-3/2}$.
\end{abstract}


\section{Introduction}
One of the most compelling problems in astronomy today is understanding
the distribution and nature of the ubiquitous dark matter, which
dominates the dynamical evolution of much of the observable universe.
Historically, much of the strongest evidence for dark matter comes
from the virial analysis of clusters of galaxies.  The high mass-to-light
ratios ($M/L \sim 300$~h) thus obtained are supported by the observed high
temperatures of the X-ray emitting gas and by the presence of giant arcs
and arclets in the clusters.  The nature and the extent of dark
matter distribution in clusters, however, is not well understood.  In the
past, information regarding the dark matter distribution in clusters could
only be extracted from dynamical studies of galaxies or from X-ray surface
brightness data, under restrictive assumptions.  For example, virial studies
generally assume that galaxy orbits are isotropic and that the light traces
the  dark matter while analyzes of the X-ray data are based on the assumptions
that the gas is in hydrostatic equilibrium within the cluster potential
and is supported against collapse by thermal pressure.  The results of
these studies can be in considerable error if these assumptions
are not valid.  Gravitational lensing, on the other hand, is a well-understood
physical process that directly probes the clusters' gravitational
potential wells.
Hence,  studies of both the strong as well as the weak
distortions in the images of faint, background galaxies induced by the
clusters offer a unique opportunity to directly probe the cluster mass
distributions in a model independent fashion, with the weak distortions being
particularly suited for mapping the mass distribution out to large radii.

Mapping the dark matter distribution in clusters using the weak gravitational
lensing effect is a subject of
great modern interest and is rapidly
becoming a mature subject.  The procedure was pioneered by Tyson \etal (1990)
and several groups have discussed techniques for acquiring and analyzing
data (Bonnet \& Mellier 1995;
Kaiser, Squires \& Broadhurst 1995; Fischer \& Tyson 1995).  The first
fully developed algorithm for constructing cluster mass maps using the weak
distortions was proposed by Kaiser \& Squires (1993).
This algorithm requires shear information extending out to infinity and
applying it to data of finite spatial extent results in a mass map with
a small but well understood
bias at the edges of the data region.  Minor modifications to the algorithm
can correct for this bias and several modified techniques have been
proposed (Kaiser \& Squires 1995;
Schneider \& Seitz 1995; Seitz, C. \& Schneider 1995;
Seitz, S. \& Schneider 1995).
We emphasize that this affects only the 2-D reconstructions of
the surface density. The aperture densitometry used here and
in Fahlman \etal (1994) is entirely free of this bias.

The list of clusters analyzed using these algorithms is growing
steadily and includes A1689 (Tyson \etal~1990; Tyson \& Fischer 1995;
Kaiser \etal 1995), Cl1409+52 (Tyson \etal 1990), MS1224+20
(Fahlman \etal 1994), Cl0024+17 (Bonnet \etal 1994; Mellier \etal 1994),
Cl1455+22 and Cl0016+16 (Smail \etal 1994, 1995).  A more complete list
appears in the review by Fort \& Mellier (1994).   All of the methods,
however, are limited in that the mass surface density can only be
determined up
to an additive constant.  Broadhurst \etal (1994) have proposed a method for
breaking this baseline degeneracy and the method was recently applied to
A1689 (Kaiser, Broadhurst \& Squires 1995).

In this paper, we present a study of A2218.  A2218 has been studied
extensively in several wavelengths.   It is an optically compact
(Butcher \etal 1983) Abell richness class 4 cluster (Abell \etal 1989) at
redshift z = 0.175.  It has a central velocity dispersion of
$1370^{+160}_{-210}$~km/s (Le Borgne \etal 1992).   A detailed photometric
and spectroscopic study of the cluster center suggests that the cluster
consists of two galaxy concentrations;  the larger of the two centered about
the sole cD galaxy in the cluster while the smaller concentration is
located $67^{\prime \prime}$
to the southeast (Pello \etal 1988; Pello \etal 1992).  Deep optical
images of the cluster have revealed a wealth of arcs and arclets,
with the arcs centered about both mass concentrations.  The location and
the morphology of the arcs suggest that the bulk of the cluster
mass is associated with the galaxy concentration surrounding the cD
galaxy.  Kneib \etal~(1995) have attempted to use the arcs to constrain the
mass distribution within the central $\sim 130$~h$^{-1}$kpc.

A2218 also exhibits a strong Sunyaev-Zel'dovich effect (Birkinshaw \etal 1981;
Birkinshaw \& Gull 1984; Partridge \etal 1987; Klein \etal 1991,
Jones \etal 1993; Birkinshaw \& Hughes 1994)
and has been well studied in the X-ray.  It has an X-ray luminosity of
$6.5 \times 10^{44}$~erg/s in the 0.5-4.5~keV band (Perrenod \& Henry 1981)
and $1\times 10^{45}$~erg/sec in the $2$--$10$~keV band
(David \etal 1993).  Its Einstein IPC image revealed a smooth circular
profile (Boynton \etal 1982).
The cluster has also been observed using both the ROSAT PSPC
(Stewart \etal 1994) and HRI instruments.   The peak of the X-ray surface
brightness distribution is coincident with the location of the cD galaxy and
the temperature of the X-ray emitting gas
has been determined to lie in the range $6$--$8$~keV
(McHardy \etal 1990; Yamashita 1995).

In a recent study, Miralda-Escud\'e \& Babul (1995) drew attention to an
interesting puzzle concerning A2218: The mass in the central regions
inferred from strong lensing distortions is greater than that determined
using the X-ray observations by a factor of $\sim 2$
if the gas is assumed to be in thermal-pressure-supported hydrostatic
equilibrium.  This led them to speculate that perhaps the gas, at least
in the central regions, may be partially supported against gravitational
collapse by means other than thermal pressure (see also
Loeb \& Mao 1994).  This mass discrepancy, if it extends out to larger
radii, has important implications for quantities, such as the cluster
gas fraction $M_{\rm gas}/M_{\rm tot}$, derived solely from X-ray data.
Typically, the cluster gas fraction is estimated to be
$M_{\rm gas}/M_{\rm tot}\sim 0.05 h^{-3/2}$ (White \& Fabian 1995;
David, Jones \& Forman 1995) a result that has been a source of much discussion
(White 1992; Babul \& Katz 1993; White \etal 1993).
This value, however, may be an overestimate by
as much as a factor of $\sim 2$ if the X-ray/lensing mass discrepancy
discussed by Miralda-Escud\'e \& Babul (1995) extends beyond the cluster core.

In this paper, we use the observed weak distortions in the images of faint
background galaxies to determine the surface mass distribution associated
with A2218 out to a distance of $\sim 600$~h$^{-1}$kpc.
We compare the distribution of total mass to that of the light and the
gas (derived from ROSAT data).  We also compare the lensing
estimate of the total projected mass in the cluster with that determined
using the X-ray data,  and estimate the mass-to-light ratio for
the cluster as well as the fraction of the total mass contributed by
baryons.  We compare the latter against constraints from nucleosynthesis.

\section{Data Acquisition}

A2218 was observed using the 3.6m Canada-France-Hawaii telescope
on the nights of 1994 June 6-9. The detector used was the 2048~x~2048
Loral 3 CCD at prime focus with a pixel size of 0\farcs207. Our observing
strategy was to take relatively short (20 minute) exposures in each band
and observe each field several times, with small random offsets.

Our I band observations of A2218 comprised of seven 20-minutes images
centered on the cD galaxy --- six of these were discarded due to bad
seeing conditions --- and two 20-minutes image of each field in a
$2\times 2$ overlapping grid about the cluster center.  The two exposures
of each field were offset by $\sim~10^{\prime \prime}$.
The observations cover a $\simeq 145$
square arcminutes field, extending to
a radius of $\simeq 6\farcm5$ ($\sim 750$~h$^{-1}$kpc) from the cluster center.
The seeing varied from FWHM=1\farcs2 during the first night to 0\farcs7 during
the second. We used only the best seeing data in our lensing analysis.
We, therefore, have a total of $9\times 20$-minutes exposures on the cluster
(one central field and $2\times 4$ grid fields).

We also observed the cluster center in V, acquiring two  20-minutes exposures
with seeing conditions of 1\farcs0 FWHM.  The V-band information, in
conjunction with the I-band data, allowed us to identify the cluster galaxy
sequence in the color-magnitude plane in order to facilitate an estimate
of the light in the cluster.  Also, the total integration of 40 minutes in the
V-band was sufficient to yield an independent mass map out to a radius of
3\farcm5 ($\sim 400$~h$^{-1}$kpc) from the cluster center.

\section{Data Analysis}

The large number of exposures that we collected over the three nights
of observing implies that in principle we ought to be able
to construct a median sky flat. Unfortunately, all of the fields
that we observed contain very luminous, extended galaxies.  As well, several
stars saturated and contaminated large portions of each exposure. The
resulting median skyflat contains ``shadows''
of these objects.  Attempts to mask the luminous galaxies and
using only frames with large offsets did not result in any significant
improvement.  Consequently,  we employed the median twilight flat.
Dividing each image  (after subtracting the bias) with the median twilight
flat resulted in an rms noise in the sky background on
the individual images of 26.8 magnitudes per square arcsecond in I and
25.6 magnitudes per square arcsecond in V.

The data was calibrated against photometric standards in the globular
clusters M92 and NGC 4147 (unpublished photometry from Davis 1990;
see also Stetson and Harris 1988 and Odewahn \etal 1992) and
Landolt (1992) standards in SA110. Color terms were found to be
unnecessary in the transformation and the I and V zero points were
determined with a formal error of less than 0.005 mag.

We identified objects on our images, and measured their position and
size, using our standard procedure (Kaiser, Squires \& Broadhurst 1995).
Briefly, we smoothed each image with a family of Gaussian filters whose
characteristic scales ranged from 0.5 to 100 pixels, varying in logarithmic
steps of $d\ln(r) = 0.2$.  The upper limit of the range
for the filter scale was chosen so that the most luminous and extended galaxies
would be comfortably contained within the filter.  We tracked the peak
trajectories (i.e.~the peak significance as a function of smoothing radius)
and assigned a radius to each object, $r_g$, corresponding to the radius of
maximum significance (i.e.~ maximum signal-to-noise).
For objects modeled as gaussian ellipsoids, this radius corresponds to
the scale length.

For each object detected, we measured the shape, magnitude, and profile
parameters. The magnitude of an object was estimated using an aperture of
radius $3r_g$ (i.e.~three times the radius of maximum significance). The
weighted quadrupole moments, $Q_{ij}$, of the surface brightness
distribution were determined using a bi-linear model for the local sky
determined in a 16-32 pixel collar around each peak. The analysis employs
a gaussian weighting function whose
scale is matched to the object's radius of maximum significance.
For gaussian objects, this choice is optimal.

The perturbation to the observed surface brightness due to a planar
lens is $f'(\theta_i) = f(\theta_i - \phi_{,ij}\theta_j)$ where
$\phi$ is the surface potential satisfying Poisson's equation
$\nabla^2 \phi = 2 \Sigma / \Sigma_c = 2 \kappa$, and
$\phi_{,ij}\equiv\partial^2\phi/\partial x_i\partial x_j$.
Here, $\Sigma_{crit}^{-1} = 4 \pi G D_l \beta$  is
the critical surface density,  $D_l$ is the angular diameter
distance to the lens and $\beta = \hbox{max}( 0, \langle 1 - w_l /w_s
\rangle )$.  In an Einstein de-Sitter universe with $\Omega = 1$,
the comoving distance $w$ is defined as $w = 1 - 1/\sqrt{1+z}$.
In the weak lensing limit, the effect of a planar lens can be represented
as a perturbation to the intrinsic quadrupole moment of an object's surface
light distribution.

We used the measured quadrupole moments to form a 2-component polarization
\begin{equation}
\epsilon_1={ Q_{11} - Q_{22} \over Q_{11} + Q_{22}}\ \ \ \ {\rm and}\ \ \ \
\epsilon_2= {2 Q_{12}\over Q_{11} + Q_{22}}.
\end{equation}
Assuming that the $\phi_{,ij}$ terms are small and constant over the
size of the galaxy image, we can relate the two polarization components
($\epsilon_1$, $\epsilon_2$) to the derivatives of the potential. For
unit weighting and intrinsically circular objects, the
polarization is related to the shear, $\gamma$, as
$\epsilon_\alpha = 2 \gamma_\alpha$ where
$\gamma_1 =  (\phi_{,11} - \phi_{,22} )/2$ and $\gamma_2=\phi_{,12}$.

For non-unit weighting (which is necessary to suppress divergent
sky noise contributions), the first order shift in polarization induced
by the shear is (Kaiser, Squires \& Broadhurst 1995)
\begin{equation}
\delta \epsilon_\alpha = P^{shear}_{\alpha \beta} \gamma_{\beta}
\label{eqn:psh}
\end{equation}
where $ P^{shear}_{\alpha \beta}$ defines the shear polarisability tensor.
This tensor can be expressed as a combination of the angular moments
of the surface brightness, the weight function and its derivatives.
Consequently,
it can be computed using quantities measured directly from each galaxy image.
The details of the calculation are involved and are described in
Kaiser, Squires \& Broadhurst (1995).  Here, we simply wish to highlight that
the observed polarization is related to the shear (and hence the lens
surface potential) in a simple linear manner.

Gravitational lensing is not the only process that can introduce
perturbations in an image's quadrupole moment.  Various other  effects,
such as guiding errors, wind shake, etc., can cause similar distortions.
Fortunately, these can be measured.  In the absence of these non-gravitational
effects, the images of the stars in the field would be
circular.  The smearing due to these effects introduces
an anisotropy in their point spread function (psf).  We model
these non-gravitational distortions as a convolution
of the actual light distribution of individual images with a small but
highly anisotropic kernel $g(\vec{\theta})$.  The calculation of the effect
in the images is very similar to the above and the resulting shift
in the polarization depends only on angular moments of $g(\vec{\theta})$.
Specifically,
\begin{equation}
\delta\epsilon_\alpha = P^{smear}_{\alpha\beta}p_\beta
\label{eqn:eshift}
\end{equation}
where $p_1 = q_{11} - q_{22}$ and $p_2=2q_{12}$
(Kaiser, Squires \& Broadhurst 1995). To first order in $p_\alpha$,
\begin{equation}
q_{ij} = \int d^2\theta \theta_i \theta_j g(\vec{\theta}).
\label{eqn:psm}
\end{equation}
The smear polarisability tensor $P^{smear}_{\alpha \beta}$ again
depends only the observed object surface brightness, the weight function and
its derivatives.

The anisotropy in the psf (characterized by $p_\alpha$) can be determined
and corrected for as follows: We isolate the stars based on a
size-magnitude cut and for each stellar image, we measure the polarization
$\epsilon_\alpha$, which in effect is a measure of the perturbation since
the actual images are expected to be circular.  Since
the smear polarisability tensor for intrinsically circular objects
is diagonal and its elements can be determined from measured quantities,
we can directly compute $p_\alpha=\epsilon_\alpha/P^{smear}_{\alpha\alpha}$
(no summation) as a function of position across the frame.  Each object's
polarization is then corrected by $ -P^{smear}_{\alpha \beta} p_\beta$,
which restores the polarization to what
would have been observed had the psf been  perfectly isotropic.

\section{Photometry}

In Figure \ref{fig:IvsR}, we plot the half-light radius $r_{hl}$
versus magnitude of all objects detected in the I-band data and in
Figure \ref{fig:VvsR}, we do the same for the V-band data.
The points enclosed within the rectangles define the stellar loci.
In constructing the catalogues,  we masked the regions in the
neighborhood of very bright stars as well as in the corners of the
fields (these were affected by vignetting), and selected objects on each
frame independently, requiring a significance of $\nu > 4 \sigma$ over
the local sky background.
The significance threshold was determined after some experimentation
with HST images. We found that this
procedure, rather than a magnitude selection, was preferable  as it
resulted in a higher signal-to-noise when measuring the shear.
We only retained objects that are detected on at least two of our
nine exposures within a tolerance of 2 pixel positional coincidence.
We used the same selection criteria to identify objects in our V-band data.
In all, we detected 4444 objects in the I-band and 1355 in the V,
corresponding to a surface
density (of galaxies and stars) of $\bar{n}_I \simeq 30$ per square
arcminute and $\bar{n}_V \simeq 28$ per square arcminute, respectively.

\begin{figure}
\myplotone{IvsR.ps}
\caption{The half-light radius vs I magnitude for all objects in our field.
We detect 4444 objects with significance $\nu > 4\sigma$ over the local
sky background for a surface density of $\bar{n}_I \simeq 30$
per square arcminute. The rectangle delineates the stellar locus.}
\label{fig:IvsR}
\end{figure}

\begin{figure}
\myplotone{VvsR.ps}
\caption{The half-light radius vs V magnitude for all objects in our field.
We detect 1355 objects with significance $\nu > 4\sigma$ over the local
sky background for a surface density of $\bar{n}_V \simeq 28$
per square arcminute. The rectangle delineates the stellar locus.}
\label{fig:VvsR}
\end{figure}

We used our catalogues to construct the I- and V-band number-magnitude
counts.  In Figures \ref{fig:Icounts} and Figure \ref{fig:Vcounts}, we
compare these counts to the I-band counts of  Lilly (1993) and  I-,
V-band counts of Woods \etal (1995). We computed the counts using all
the objects (solid histogram) and after eliminating all objects that lie
on the stellar locus in Figures \ref{fig:IvsR} and \ref{fig:VvsR}
(dashed histogram).  The counts are complete to $I \simeq 23$ and
$V \simeq 24$. As the plots show, the counts in our frames are elevated
with respect to the field counts, particularly at bright magnitudes.
Removing the stars does not eliminate this
discrepancy.  The excess is mainly due to contamination by cluster galaxies,
which we have not attempted to remove.  The agreement with the field counts
improves towards fainter magnitudes.

\begin{figure}
\myplotone{Icounts.ps}
\caption{The counts per square degree per magnitude from the I-band
data.  The solid histogram is using all objects; the dashed histogram
comes from removing bright stars ($I<21.5$).
The solid line is a fit to the Woods \etal (1995) counts.
At the brighter magnitudes,
we see an excess in the counts due to the cluster galaxies. At $I\simeq 22$
the counts agree reasonably well with the field counts. The
sample is complete to $I \simeq 23$.}
\label{fig:Icounts}
\end{figure}

\begin{figure}
\myplotone{Vcounts.ps}
\caption{The counts per square degree per magnitude from the V-band
data.  The solid histogram is using all objects; the dashed histogram
comes from removing bright stars ($V<22.5$). The solid line is a fit to the
Woods \etal (1995) field counts.}
\label{fig:Vcounts}
\end{figure}

In Figure \ref{fig:ColourvsI}, we display the color-magnitude diagram for
the central field. A red sequence of objects with a mean color of
$ V-I \simeq 1.6$ at the bright end is clearly distinguishable.  We
identified these objects as cluster galaxies. To extract this
sample, we fitted a linear model to the color sequence and select
objects with color within 0.2 magnitudes of the mean.  In Figure
\ref{fig:ClusterColPlotI}, we plot the I-band number counts for the
color selected sample of cluster galaxies.  The sample contains
196 galaxies with $I< 20.25$.  The number of objects per magnitude is
roughly constant.  Fitting the cluster luminosity function with a
Schecter function $n(m) = A 10^{-0.4 m ( \alpha + 1 )}$ yields
$\alpha = -1$ for the faint end slope.

\begin{figure}
\myplotone{ColourvsI.ps}
\caption{The V-I color vs I for the center field.}
\label{fig:ColourvsI}
\end{figure}

\begin{figure}
\myplotone{ClusterCountsI.ps}
\caption{The counts per square degree per I magnitude for objects with
$I<20.25$
and color $\pm 0.2$ from a linear fit to the color sequence. The dashed line
shows the Woods \etal (1995) field counts.}
\label{fig:ClusterColPlotI}
\end{figure}

\section{Light Distribution}

We computed the galaxy surface number
density distribution as well as the corresponding light
distribution using all of the galaxies in the images.
To reject obvious foreground galaxies, we applied a very conservative
mask. We removed the bright spiral galaxy to the northeast
of the cD galaxy but allowed its companion to remain. The giant elliptical
to the east is very likely to be a foreground object as well. We
allowed it to remain in the calculation of the light distribution
however, as the is a significant overdensity of galaxies in its vicinity
suggesting that there is perhaps a group or small cluster associated with
the galaxy.

We display  the light and galaxy surface number density as contour plots
superposed on the optical image of the cluster
in Figures \ref{fig:numberonimage} and \ref{fig:lightcontour}, respectively.
The contours have been smoothed with a Gaussian smoothing scale of 0\farcm67
and are broadly similar. The peak of the light map
(Figure \ref{fig:lightcontour})
is centered on the primary galaxy clump (the one containing the cD galaxy)
in the cluster and the surface brightness drops off smoothly towards the
cluster periphery. In the central region, the contours are elliptical,
with the major axis aligned with the primary and secondary galaxy
concentrations as well as the chain of bright galaxies stretching away from
the cD galaxy in the northwest direction. At distances greater than
$\sim 2^\prime$ from the cD galaxy, the contours show a weak extension
towards a giant elliptical galaxy to the east of the cD galaxy. The surface
number density plot (Figure \ref{fig:numberonimage}), has a very similar
morphology. The peak and orientation of ellipticity matches that of the
light map. The extension to the east is more evident and extends beyond
the giant elliptical galaxy. In addition,
there are extensions roughly to the north, south, and west/northwest.
Employing the color selected bright galaxy sample in place of the entire
galaxy sample yields a very similar galaxy number density and light
distribution. The peak and central ellipticity match that of the entire
galaxy population in the images. There is some weak evidence of the
extension towards the east, but it is not well resolved in the color
selected sample.

\begin{figure}
\myplotfiddle{numberonimage.ps}
\caption{The smoothed surface number density of galaxies (contours) placed
on the optical I-band image of the cluster center. The contours
have been smoothed with a gaussian of scale length 0\farcm67. North is
to the right; East is up.}
\label{fig:numberonimage}
\end{figure}

\begin{figure}
\myplotfiddle{lightcontour.ps}
\caption{The smoothed light from all the galaxies (contour) superimposed on
the optical image of the cluster field. The light
contours have been smoothed with a gaussian of scale length 0\farcm67. North is
to the right; East is up.}
\label{fig:lightcontour}
\end{figure}

While qualitative descriptions of the light and galaxy distributions are
useful, to understand the nature of the distributions and to make comparisons
with the distributions of gas and total mass in the cluster, we need to be
more quantitative.  The usual approach is to compute the mass-to-light ratio
as a function of radius.  In principle, we can use the excess in the number
of galaxies over the field counts to estimate the number of galaxies
in the cluster and hence, estimate the cluster light;  however, we prefer
a more simple approach.  We make two measurements, which place upper and
lower bounds on the cluster light.

To determine the upper bound, we estimate the light contributed from
{\it all} galaxies in our observations.  This sample will certainly contain
many field galaxies and hence yields an
overestimate of the light contributed by the cluster galaxies.  We work
directly with the co-added V image and compute a cumulative, circularly
averaged, radial light profile centered on the cD galaxy.  We select all
objects that were detected on both of the V-band
images and identify as galaxies all objects having half-light radius
greater than 1.2 times the mean stellar half-light radius. We mask all the
galaxies on the image and determine a light profile of the resulting image.
The light profile of the galaxies is then determined by
subtracting the profile of the masked image from that of the unmasked image.

To establish the lower bound, we do a similar analysis using only our
bright color selected galaxy sample.  At the bright end, we do not expect
much contamination from the field galaxies;  most
galaxies ought to be cluster members.  The color selected sample, however,
excludes all faint ($I>20.25$) cluster
galaxies as well as bright cluster galaxies that lie outside the
color sequence. Hence, the resulting cumulative light profile
corresponds to a lower bound on the total cluster light.

The two cumulative light profiles described above are shown in
Figure \ref{fig:cumVvsR}.
The solid line corresponds to the light profile computed using all the
galaxies and the short-dashed line corresponds to that computed
using only the bright, red cluster galaxies.
At a fiducial radius of 3\farcm5 $(\sim 400$~h$^{-1}$kpc),
the total apparent magnitude is
$V = 13.68$ while that computed using only the red galaxies is
V = 13.96.  These two estimates
place an upper and lower bound on the cumulative magnitude at this radius.
We determine also the  cumulative V-band profile
using all galaxies in the more spatially extended I-band images, converting
from I to V by applying a color transformation
of $V-I = 1.6$, which matches the bright end of the color sequence.
In the region where the the two profiles overlap, the latter profile
(the long-dashed line in Figure \ref{fig:cumVvsR})
agrees very well with that computed directly using the V-band data.

It is customary to quote the luminosity (or the luminosity profile)
in the B-band.  We estimate the B-band luminosity for A2218 using our
apparent V-magnitudes as
\begin{equation}
L_B = 10^{ 0.4 ( M_{B \odot} - V - ( B - V ) + DM + K) }L_{\odot}
\end{equation}
where $M_{B \odot} = 5.48$ is the total solar B magnitude. We apply
a K-correction of $K = 0.3$ and a color transformation of $B-V = 0.93$,
as suggested by the data in Coleman, Wu and Weedman (1980).  Since A2218
is at redshift $z=0.175$, its distance modulus is $DM = 38.69$ (for
$\Omega = 1$ and $h = 1$).

The cumulative B-band luminosity, as a function of radius, is plotted in
Figure \ref{fig:cumLvsR}.  Within the central $\sim 100^{\prime \prime}$,
the profile agrees reasonably well with that of Kneib \etal~(1995).
At larger radii, the cumulative light in the cluster increases
linearly with $r$, suggesting that the optical surface brightness varies
with radius as $1/r$.  At a fiducial radius of
3\farcm5 ($\sim 400$~h$^{-1}$kpc), the estimated  B-band luminosity of
the cluster is in the range  $6.8$--$8.8\times10^{11}$h$^{-2} L_\odot$.

\begin{figure}
\myplotone{cumM.ps}
\caption{The cumulative V magnitude as a function of radius from the cD
galaxy. The solid line comes from the V-image; the short-dashed line
is the red cluster galaxy contribution to  the light. The long dashed line is
the calculation done on the I-band image with a transformation to V being
done with a constant color shift of $V-I = 1.6$.}
\label{fig:cumVvsR}
\end{figure}

\begin{figure}
\myplotone{cumL.ps}
\caption{The cumulative luminosity as a function of radius from the cD
galaxy. The lines are as described on Figure \protect{\ref{fig:cumVvsR}}.}
\label{fig:cumLvsR}
\end{figure}

\section{Lensing Analysis}

As discussed in \S 3, we used the stellar psf to determine and correct for the
distortions introduced by non-gravitational influences in the images of all
objects on a given frame (see equations \ref{eqn:eshift} and \ref{eqn:psm}).
We found that the spatial variations in the psf, over a given image frame,
is well modeled by a second-order polynomial.  The constant term in the
polynomial is benign for our lensing analysis -- a uniform
shear does not affect our reconstructions of the surface mass density.
However, the gradient and the higher order terms are important and
contaminations affecting these quantities are more worrisome.
In the uncorrected images, we found a gradient in the stellar psf that
corresponded to a shift in polarization of 4\% across each frame.  In the
cluster outskirts, this is approximately equal to the
signal we are trying to measure, and so it clearly must be removed.
In addition, we also found that the distortions in the stellar psf,
particularly the values of the second-order terms in that
expansion, varied from frame to frame.  This not too surprising and is
likely due to the unintended rotation of the bonette between exposures.
These second order terms, however, introduce only a small
correction and their inclusion in the analysis resulted in only a very slight
improvement over the gradient
model, as measured by the change in the residual chi-square value.

We show the psf for one of our frames in Figure \ref{fig:stellarpsf}. The raw
frame (left panel) has a mean $\epsilon_1 = -0.07$. The right panel shows the
polarization after the correction is applied. In this case, we are left with
a residual mean polarization of $0.0008$. In all of our corrected frames,
the residual polarization is $\leq 0.005$.

\begin{figure}
\myplotone{stellarpsfcorr.ps}
\caption{The point spread function for one of the I-band images. The left
panel shows the major axis of objects identified as stars in the field. The
right panel shows the major axis of the stars after the correction has been
applied. We use a frame dependent, second order model
for the psf. Here the raw frame had a net $\epsilon_1 = -0.07$, while after
the correction is applied the residual $\epsilon_1 = 0.0008$. The residual
polarization was $< 0.5$\% for all of our corrected frames.}
\label{fig:stellarpsf}
\end{figure}

As discussed previously, we identified as galaxy candidates those objects, on
each corrected frame, that appear on at least two frames and have a
half-light radius  at least 1.2 times larger than the stellar
psf ($r_{hl} > 0\farcs26$).  We then
designated those galaxies with magnitudes  $I>21$ as background objects.
This yielded a catalogue containing 2370 background galaxies over an area of
145 square arcminutes around the cluster center.  We converted the
individual polarization estimates to shear
measurements according to equation \ref{eqn:psh}.

Next, we calibrated the effect of atmospheric seeing, which tends to
diminish the shear. We used the Medium Deep Survey (MDS) data from the
Hubble Space telescope to accomplish this.
The raw HST images have roughly twice the resolution of our CFHT data and are
unaffected by the atmosphere.  The combination of the two data sets allowed
us to  quantify the damping of the shear due to atmospheric smearing.
Using the WFPC2 fields, we constructed a mosaic that is twice the
size of our survey field and applied a constant shear of amplitude
$\gamma = 0.15$. The latter simulates well the effect of lensing
on the galaxy population by a low redshift cluster.  We then rebinned
the pixel size to match the resolution
of our CFHT data, and applied a gaussian smoothing to simulate the observed
seeing conditions. Strictly, this should slightly underestimate the dilution
of the measured shear due to seeing as the observed psf has a small extended
wing compared with a gaussian. However since we compute centrally weighted
quadrupole moments, the slight extension of the psf contributes a minimal
amount and and the results presented here do not change if we use the shape
of the psf measured from our data.
Finally, we added sky noise comparable to that in our
observations.  The resulting image was processed in
exactly the same manner as the real data.  Since we are using real galaxy
images as input and are simulating, as nearly as possible, the observing
conditions we had for our program,  we are able
to empirically calibrate the loss in the shear signal.  In this fashion,
we also do not need to model the intrinsic distributions of galaxy
sizes, magnitudes and ellipticities.

Subjecting the simulated CFHT images to the same significance, size and
magnitude cuts as the data, we found that for the I-band data we were able
to recover $(50 \pm 7)$\% of the input
signal, with the dispersion being estimated from the scatter over
several simulations. All of our shear estimates are boosted by this factor.
In Figure \ref{fig:etonlight}, we display the mean shear calculated
on a $16 \times 16$ grid placed on the I-band image of the field
about A2218. The grid calculation
ensures that all points are strictly independent. The visual pattern
is suggestive and the tangential alignment around the cD galaxy is evident.

\begin{figure}
\myplotfiddle{etonlight.ps}
\caption{The mean shear calculated on a $16 \times 16$ grid placed on the
I-band image of the field. The length of the vector is proportional to
the shear with the longest line being equivalent to a shear
of 65\%.}
\label{fig:etonlight}
\end{figure}

In the upper panel of Figure \ref{fig:etprofI}, we plot the mean radial
tangential shear profile, using the cD galaxy as the center.  The upper
dashed line shows the prediction for a
singular isothermal sphere with the observed velocity dispersion
of 1370~km/s (Le Borgne \etal 1992) , while the lower line corresponds to
$\sigma = 1000$~km/s.  In the lower panel, we display the statistic
(Kaiser \etal 1995)
\begin{eqnarray}
\zeta(\theta_1, \theta_2)& =& 2( 1 - \theta_1^2 / \theta_2^2 )^{-1}
	\int_{\theta_1}^{\theta_2} d \ln(\theta) \langle \gamma_t \rangle \\
  & = & \bar{\kappa}(\theta_1) - \bar{\kappa}(\theta_1 < \theta < \theta_2)
	\nonumber \\
\end{eqnarray}
which measures the mean surface density interior to $\theta_1$ relative
to the mean in an annulus $\theta_1 < \theta < \theta_2$. The dashed
lines show the predictions for the two isothermal sphere models assuming
that the cluster extends into the control annulus.

We repeated this analysis for our V-band data and display the corresponding
radially averaged tangential shear profile and the $\zeta$ statistic in
Figure \ref{fig:etprofV}.
The similarity between the V- and I-band results is reassuring; the fact that
the measured shear estimates are reproducible in different wavelengths implies
that we are not introducing some wavelength-dependent bias in our analysis.

\begin{figure}
\myplotone{etprofI.ps}
\caption{The radially averaged tangential shear from the I-band data.
The dashed line are predictions of the isothermal sphere model with
$\sigma = 1000, 1370$ km/s respectively. The bottom panel shows the
statistic which measures the mean $\kappa$ interior to $\theta$, relative
to the mean $\kappa$ in the annulus $\theta < \theta' < 373^{\prime \prime}$.
The error bars are estimated from the orthogonal shear component - this will
overestimate the error if the shear pattern is non-circular.}
\label{fig:etprofI}
\end{figure}

\begin{figure}
\myplotone{etprofV.ps}
\caption{The radially averaged tangential shear from the V-band data.
The solid lines are the isothermal sphere models with
$\sigma = 1000, 1370$ km/s. The errors are calculated from the second
moment of the orthogonal shear component.}
\label{fig:etprofV}
\end{figure}

The plots in Figures \ref{fig:etprofI} and \ref{fig:etprofV} show that
for $r>180^{\prime \prime}$ ($r>340$~h$^{-1}$kpc), the measured tangential
shear is consistent with that expected from a singular isothermal
mass distribution with the observed velocity dispersion
($\sigma = 1370$~km/s). Interior to this, the measured shear more
closely corresponds to that expected
from a $\sigma=1000$~km/s isothermal mass distribution. One can interpret
this trend as implying that the velocity dispersion characterizing the
mass distribution in the cluster is $\sigma \sim 1000$~km/s out to a
radius of $r=100^{\prime \prime}$ ($r=190$~h$^{-1}$kpc) from the cD
galaxy and then, smoothly increases to $\sigma \sim 1370$~km/s
at larger radii.  Alternatively, the cluster velocity dispersion may very
well be constant throughout the cluster but the observed value of shear
in the central region is suppressed because the sample of galaxies used
to compute the shear is strongly contaminated with cluster
galaxies.
Certainly, we can get an indication that shear suppression due to cluster
contamination is indeed happening by focusing only on those galaxies
with colors redder than the main cluster sequence.  By and large, these
galaxies will be background objects.  The resulting mean tangential shear
at small radii is greater than that determined using all galaxies.
Unfortunately, the number of red background galaxies is small and
hence, we are unable to offer anything more definitive.
In any case, our conversion of the measured quadrupole moments of
the galaxy image into shear is based on the assumption that the
derivatives $\phi_{,ij}$ of the surface potential are small.
This assumption certainly breaks down in the central regions of the cluster.

At large radii, the cluster contamination is not an important consideration.
Each of the points in Figure \ref{fig:etprofI} with $r > 180^{\prime \prime}$
corresponds to an average over $\geq 300$ galaxies.  Furthermore, the
measured shear is small (of the order of 10\%) and hence, we are
confident about the results yielded by our analysis.

Turning back to the actual distribution (as opposed to the
radial profile) of shear across cluster, we used the maximum
probability extension of the original inversion algorithm
(Kaiser \& Squires 1993; Kaiser \& Squires 1995) to construct a map of
the dimensionless surface (total) mass density, $\kappa$.  This modified
algorithm takes into account the finite nature
of the data and yields an unbiased estimator of the surface mass
density with very low noise properties.
In Figure \ref{fig:massonI}, we superpose the contour map of the
reconstructed $\kappa$ on the optical image of the cluster. The contour
maps has been smoothed with a gaussian of scale length
0\farcm67.  The location of the dominant peak of coincides with the
position of the cluster cD galaxy with a secondary peak
approximately  $3^\prime$ away towards the east.
There is, however, no resolved peak
associated with the secondary galaxy concentration although the isodensity
contours in the  vicinity of the cD galaxy are elongated towards
the secondary concentration.
Away from the cD, the isodensity contours show extensions towards
the north, west, east and
south-by-southeast, with the eastern extension forming a bridge between
the two peaks.

In order to convert $\kappa$, the dimensionless surface density, into
a physical quantity, we need to determine the critical surface density
$\Sigma_{crit} = (4\pi G D_l \beta)^{-1}$.
For low-redshift clusters, the value of $\Sigma_{crit}$ is only weakly
dependent on the background galaxy redshift distribution
(Kaiser, Squires, \&
Broadhurst 1995).  We adopted a value of $\beta = 0.6$ as suggested by
an extrapolation of the observed redshift distribution (Lilly 1993, 1995;
Tresse \etal 1993) to fainter magnitudes and hence,
$\Sigma_{crit} = (7.0 \pm 0.7) \times 10^{15}$~h~M$_\odot$/Mpc$^2$,
where we assumed a 10\% uncertainty in $\beta$.
We note that the  value of $\beta$ is not very sensitive to the model
used to estimate the redshift distribution; for example, the model for
the faint galaxies of Gronwall \& Koo (1994) yields $\beta = 0.59$.

\begin{figure}
\myplotfiddle{massonI.ps}
\caption{The reconstructed surface density using the maximum probability
extension to the original Kaiser-Squires algorithm. The contours have been
smoothed with a gaussian with scale length 0\farcm67.}
\label{fig:massonI}
\end{figure}

In Figure \ref{fig:MassVsR}, we plot the 2D mass profile of the cluster.
The solid line is the mass profile of a singular isothermal model with
a velocity dispersion of 1370~km/s and whose mass profile
extends into the control annulus.
We also plot the projected mass at $r=21^{\prime\prime}$
($40$~h$^{-1}$kpc), the presumed critical radius determined from one of the
giant arcs with redshift $z = 0.7$ (see Miralda-Escud\'e \& Babul
1995), as well as the projected mass at
$r=68^{\prime\prime}$ ($130$~h$^{-1}$kpc)
implied by the mass model proposed by Kneib \etal (1995) in order to
account for the positions and the morphologies of the giant arcs.
The mass estimate of
Kneib \etal (1995) is in reasonable agreement with the $\sigma=1370$~km/s
isothermal prediction at a radius of $\simeq 1^\prime$ --- their mass estimate
actually corresponds to a singular isothermal with a velocity dispersion of
$\sigma = 1200$~km/s --- but is higher than the weak lensing result.  This is
not surprising since the the weak lensing mass estimates at small radii are
expected to underestimate the true mass, as we have already discussed.
At radii where we have confidence in the weak lensing results ($r >
180^{\prime\prime}$),  the projected mass profile is consistent with that
of the $\sigma=1370$~km/s isothermal model (i.e.~$\Sigma_{tot}\sim 1/r$)
although there is a slight indication that the mass profile might be slightly
steeper.

\begin{figure}
\myplotone{MassVsR.ps}
\caption{The 2D mass profile of the cluster. The points with errors are the
estimates from the weak lensing with the uncertainties calculated from adding
the errors in $\kappa$ and $\Sigma_{crit}$ in quadrature.
The mass calculation uses the
surface density interior to a radius relative to that in the control region
extending to $\theta = 370^{\prime \prime}$.
The solid and dashed lines are isothermal sphere predictions for
$\sigma = 1370, 1000$~km/s respectively, assuming the
cluster extends into the control annulus.
The {\bf X} represents the mass
estimated by Kneib \etal 1995 by modeling the giant arcs.
The triangle is the mass measured by Miralda-Escud\'e \& Babul (1995)
at the critical radius from the cD galaxy.}
\label{fig:MassVsR}
\end{figure}

\section{X-ray Analysis}

Abell 2218 was observed by ROSAT both with the PSPC (position sensitive
proportional counter) on 1991 May 25-26 for 44530~seconds and with the
HRI (high resolution imager) on  1994 January 5-7 and  1994 June 17-19
for a total 35809~seconds. The X-ray emission originates from the hot
intracluster medium (ICM) contained within the gravitational potential of
the cluster.  A contour plot of the HRI X-ray surface brightness,
superposed on the optical image of the cluster, is presented in
Figure \ref{fig:xrayonimage}.  The X-ray map has been smoothed with a
gaussian filter of scale 0\farcm{23}.  The X-ray contours are elliptical and
the position angle of the major axis varies with distance from the central
peak.  This peak is coincident with the cD galaxy and there is
no evidence of enhanced emission (a secondary peak) associated with the
second of the two galaxy concentrations (see also Kneib \etal 1995)
although  at distances of 1\farcm5 to $2^\prime$ from the
central peak, the X-ray contours show extensions in the southeast (encompassing
the secondary mass concentration), north and west/west-by-northwest
directions.

\begin{figure}
\myplotfiddle{xrayonimage.ps}
\caption{The HRI X-ray surface brightness profile placed on the optical
image of the cluster. The contours have been smoothed with a gaussian
with scale length 0\farcm23. North is to the right; East is up.}
\label{fig:xrayonimage}
\end{figure}

The radial profile of the observed X-ray surface brightness of A2218 (see
Figure 2 of Miralda-Escud\'e \& Babul 1995) can be well fit by a function
of the form
\begin{equation}
S(r) = S_0\left[1+\frac{r^2}{r_c^2}\right]^{-3\beta+1/2}
\end{equation}
where $S_0$ is the central surface brightness and $r_c$ is the core radius
(eg. Fabricant \& Gorenstein 1983).  To minimize contamination by the
background, we only used the PSPC data in the 0.5 to 2~keV band  and fitted
the radial profiles for each of
the four quadrants separately.  The resulting  values of $r_c$ and $\beta$
are presented in Table 1.  To account for the possible smoothing of the X-ray
data by the point spread function of the detector, we applied Lucy's
deconvolution algorithm (Lucy 1974) to the image and performed fits to the
resulting surface brightness profiles.  The difference between values of
$\beta$ and $r_c$ for the original and the deconvolved images is
insignificant. As the dispersion in $r_c$ is small, the fits show that
X-ray surface brightness can be reasonably approximated as being
circularly symmetric, with errors due to deviations from circular
symmetry being relatively small.

The above expression for the X-ray surface brightness profile can be
readily inverted to yield the gas density profile
\begin{equation}
\rho(r) = \rho_0\left[1+\frac{r^2}{r_c^2}\right]^{-3\beta/2}.
\end{equation}
The central density $\rho_0$ is proportional to the central electron
density $n_{e0}$, which is given by
\begin{equation}
n_{e0} = \frac{1.2~S_0(1+z)^4}{4 \pi \Lambda r_c}
\frac{\Gamma(3\beta)}{\Gamma(1/2)\Gamma(3\beta-1/2)}
\end{equation}
where $S_0$ is expressed in terms of arcmin$^{-2}$, and
$\Lambda$ is the emissivity of the gas in the observed energy band.
Formally, the above inversion assumes that the gas is isothermal and
spherically distributed.  However, for gas at temperatures of $2$--$15$~keV,
the photon count rate in the 0.5 to 2~keV PSPC energy band is relatively
insensitive to temperature variations; for example, the values of $n_{e0}$
computed assuming a gas temperature
of $T=3$~keV and $T=8$~keV differ only by 5 to 10\%.  The values of $n_{e0}$
associated with the fits to the X-ray surface brightness profiles
in each of the four quadrants are given in Table 1.  At large radii, the gas
surface density falls of as $\sim 1/r$ and since the X-ray surface brightness
profile extends out to a projected distance of at least
$\sim 9^\prime$ ($\sim 1$~h$^{-1}$Mpc), so must the gas distribution.

Integrating the radial gas density distribution yields the total
gas mass profile for the cluster. In Figure \ref{fig:logMassVsR}, we
show the projected (2D) gas mass profile.
The error bars reflect the uncertainties in the values
of parameters ($\beta$, $r_c$)
characterizing the surface brightness profile as well as in the
value of the gas temperature at each radius.  At the fiducial
radius of 3\farcm5 ($\sim 400$~h$^{-1}$kpc),
the projected gas mass is $(1.65 \pm 0.5)  \times 10^{13}$~h$^{-5/2} M_\odot$.
Unlike the weak lensing mass estimate, the gas mass corresponds to
the {\it actual} projected mass within a specified radius.

\begin{table}
\begin{center}
\begin{tabular}{|l|ccc|}
\hline
direction & r$_c$ in [kpc/h] & $\beta$ & n$_{e0}$ in [cm$^{-3}$]\\
\hline
\hline
west & 153 & 0.73 &  0.0068 \\ \cline{2-4}
     & 110 & 0.62 &  0.0076 \\
\hline
south & 144 & 0.78 & 0.0086 \\ \cline{2-4}
 & 110 & 0.67 & 0.0093 \\
\hline
east & 206 & 0.78 & 0.0064 \\ \cline{2-4}
 & 158 & 0.68 & 0.0069 \\
\hline
north & 138 & 0.75 &  0.0079 \\ \cline{2-4}
 & 86  & 0.63 &  0.0092 \\
\hline
\end{tabular}
\end{center}
\caption
{The results of the isothermal $\beta$-fits in the four directions from the
hard image derived from the PSPC data (we used the PSPC data for fitting the
models, as the statistics in that data set
is much better than for the HRI data, due to a lower background.)
The two different results for each direction correspond to 3$\sigma$-errors.
$n_{e0}$ is the central electron number density for
8~keV. The electron density is lowered by about 5\% to 10\% assuming a gas
temperature of 3~keV.}
\label{table:xraydata}
\end{table}

Under the assumption that the intracluster gas is in thermal-pressure-supported
hydrostatic equilibrium in the cluster potential, the X-ray data can also be
used to estimate the total gravitational mass of the cluster:
\begin{equation}
M(<r) = -\frac{kT(r)r}{G\mu m_p} \left[ \frac{d\ln\rho_g}{d\ln r}+
\frac{d\ln T}{d\ln r}\right]. \label{eq:n3}
\end{equation}
This calculation, however, requires the knowledge of the temperature
distribution in the cluster.  The first ASCA observations give
a temperature of
T=8~keV (Yamashita 1995) while GINGA observations yield an
overall flux-weighted ICM temperature of
6.7$^{+0.5}_{-0.4}$~keV (McHardy \etal 1990). A spectral analysis of the
ROSAT/PSPC data, on the other hand, yields temperatures in the range
of 3 to 5~keV.   In this analysis, we simultaneously fitted the
temperature, the hydrogen column density and the metallicity in
different annuli. The temperature determinations for the different radial
bins agree within the error bars and are always lower than the GINGA or
the ASCA determinations.

The discrepancy between our ROSAT-based determination of the gas
temperature in A2218 and the temperature measurements from ASCA or GINGA may
simply reflect the fact that the analysis of
the ROSAT data does not provide as accurate results as the other two
satellites in cases where the peak of the ICM energy spectrum lies beyond
the energy range of the PSPC.  Alternatively,
the discrepancy in the temperature determinations may indicate the presence of
gas components with different temperatures, with the PSPC data being
dominated by emission from the colder components
while the ASCA and the GINGA data is dominated by the high temperature
component.  Indeed, a new analysis of the ASCA data suggests that
the gas temperature declines away from the cluster centre (Mushotsky 1995).
We did attempt a two temperature fit to the X-ray spectra but
did not find any improvement.

As noted above, our spectral analysis also yields a measure
of the hydrogen column density and metallicity.  We find a hydrogen
column density in the range 1-3$\times10^{20}$cm$^{-2}$ and
a metallicity of 0.2$^{+0.2}_{-0.2} \,Z_\odot$.  The metallicity
agrees very well with the
GINGA and the ASCA results of 0.2$^{+0.2}_{-0.2}  \,Z_\odot$
(McHardy \etal 1990) and 0.18$\,Z_\odot$ (Yamashita 1995), respectively.

It should be noted that the results of the spectral analysis
depend weakly on the adopted value for the background.  In our analysis,
we used values estimated from several different regions.  In addition,
the temperature and the hydrogen column density determinations
also depend weakly on the data channels used.  Ignoring the lower channels
(channel 20-40; channels lower than 20 are always neglected because of the
low efficiency of the ROSAT/PSPC in that energy range) increases the
hydrogen column density and the temperature slightly.

In estimating the cluster mass from the X-ray data, we
need to take into account the possibility of
temperature gradients existing in the cluster as well as the
uncertainty  in the value of the overall
ICM temperature.  Indeed, a recent re-analysis of the ASCA data suggests that
there is a significant temperature gradient, with the new estimate of the
central temperature in good agreement with the
GINGA result (Mushotsky 1995). As the mass estimates depend both on the
radial temperature and its gradient (see equation \ref{eq:n3}),
it is important to allow for variations away from isothermality.
We accomplish this by using the Monte Carlo scheme of Neumann \&
B\"ohringer (1995) to determine the total mass profile.  Briefly, we
generate a temperature profile for the cluster by assigning a temperature,
randomly drawn from a pre-defined range (3 to 8~keV in the present case),
at equally spaced (60~h$^{-1}$kpc) locations along the radial
direction.  To prevent unrealistic short-scale oscillations in the temperature
profile, we require the temperatures at adjacent locations to be within
0.6~keV of each other.
At every step, we calculate the cumulative total mass profile and use only
those temperature profiles that give a mass profile that increases
monotonically with radius. To obtain an optimal estimate of the mean mass
profile as well as the standard deviation,
we generate $10^5$ temperature profiles and compute the corresponding mass
profiles.  To ensure that our mass determination includes
uncertainties in the gas density profile,
we vary the values of ($\beta$, $r_c$) over the range
given in Table \ref{table:xraydata} and repeat
the above operations.  The projected (2D) profile of the total
mass derived from the X-ray data
is shown in Figure \ref{fig:logMassVsR}. At the fiducial
radius of 3\farcm5 ($\sim 400$~h$^{-1}$kpc),
the projected total mass is $(2.6 \pm 1.6) \times 10^{14}$~h$^{-1} M_\odot$.
As in the case of the gas mass estimates, the projected X-ray mass estimate,
unlike the weak lensing estimate, corresponds to the {\it actual}
projected mass within
a specified radius.

Combining the above estimates of the gas mass and the total mass determined
from the X-ray data yields a gas-to-total mass ratio of
$M_{gas} / M_{tot} = (0.06 \pm 0.04)$~h$^{-3/2}$.  This is consistent with
the value derived for the Virgo cluster (B\"ohringer \etal 1994) and a
typical value for clusters using the X-ray estimates for the gas and total
mass is
$M_{gas} / M_{tot}\sim 0.05$~h$^{-3/2}$
(White \& Fabian 1995; David, Jones \& Forman 1995).

\begin{figure}
\myplotone{logMplot.ps}
\caption{The 2D mass profile of the cluster. The solid squares are
estimates from the weak lensing. The x's are total mass determined
from the X-ray gas, with the errorbars being the 2$\sigma$ dispersion
calculated from the $10^5$ simulations. The open triangles are
the corresponding estimates of the projected gas mass. The solid line is
the isothermal model with a velocity dispersion of 1370 km/s and
the {\bf X} represents the mass
computed by Kneib \etal (1995) by modeling the giant arcs.}
\label{fig:logMassVsR}
\end{figure}

\section{Discussion and Conclusions}

In this paper, we have reported on the detection of dark matter in
the cluster of galaxies A2218. We found a strong, coherent shear pattern
over a scale of $\sim 400^{\prime \prime}$ from the cluster
optical center.  Using the weak lensing technique, we have mapped out
the dark matter distribution.  We have also analyzed the
distribution of galaxies, of optical light and of the X-ray emission
associated with the cluster.

The combination of the optical, lensing and the X-ray data
affords us an unprecedented opportunity to compare the relative
distributions of the galaxies, the gas and the total mass in A2218.
These distributions are displayed in Figures \ref{fig:numberonimage},
\ref{fig:lightcontour}, \ref{fig:massonI} and \ref{fig:xrayonimage}
respectively.  In all of the distributions, the location of the peak
(or the dominant peak) corresponds to the location of the central
cD galaxy/primary galaxy concentration.  There are no strong distinct
features associated with the secondary galaxy concentration.  In the case
of the the light and the mass maps, it can be argued that the lack of such
features may be a consequence of the poor resolution resulting from smoothing
the maps on scale of 0\farcm67.  The X-ray map, however, is smoothed on the
scale of 0\farcm{23} and hence, has a much higher resolution.

The contours immediately surrounding the central peak in both the light and
the mass maps are elliptical,  with the elongation extending in
the direction of secondary galaxy clump.  The X-ray contours in the immediate
vicinity of the peak are also elliptical but with the major axis oriented
orthogonal to that of the mass and the light isocontours.
The relative orientation of the isocontours in the mass and
the X-ray maps are reminiscent of features seen in hydrodynamical simulations
of a cluster undergoing a merger with a subcluster (eg. Schindler \&
M\"uller 1993). In such events, the X-ray emission tends to trace the
lenticular shocks that expand in the direction perpendicular to the
trajectory of the infalling subcluster.  If this is indeed the case in A2218,
then the gas in the central regions would not be expected to be in
hydrostatic equilibrium and this may help explain the discrepancy between
lensing and the X-ray mass estimates reported by Miralda-Escud\'e and Babul
(1995). However, smoothing the X-ray map on the same scale as the mass and
the light maps yields an almost circular surface brightness
profile and consequently, it is difficult to ascertain whether both the
ellipticity and the orientation of the X-ray contours on scales smaller
than $\sim 1^\prime$ are significant.

At distances of $1^\prime$--$3^\prime$ from the central peak, the extensions in
the isodensity contours in the mass map closely follow those in the galaxy
surface density plot and to a lesser degree, in the galaxy light distribution.
The gas distribution as delineated by the X-ray surface brightness also
appears to trace the mass closely.  The  X-ray map exhibits all of the
extensions in the mass map with the exception of the one, the eastern
extension.

At projected distances $3^{\prime}$ ($\sim 340$~h$^{-1}$kpc) to $5^\prime$
($\sim 600$~h$^{-1}$kpc) from the cD galaxy, the radial surface density
profiles of the cluster light, the intracluster gas and the total mass are
all consistent with a $\sim 1/r$ decrease.  On these scales, there
is no indication of the light or the gas distributions being biased with
respect to the total mass.

The projected total mass within a fiducial radius of 3\farcm5
($\sim 400$~h$^{-1}$kpc)
from the cD galaxy, as yielded by the weak lensing analysis, has a lower
bound of $M_{tot} = (3.9 \pm 0.7) \times 10^{14}$~h$^{-1} M_{\odot}$.
This value is in good
agreement with the projected total cluster mass inferred from the X-ray
data under the assumption that the gas is in thermal pressure-supported
hydrostatic equilibrium.
Since the projected gas mass interior to $\sim 400$~h$^{-1}$kpc
is $(1.65 \pm 0.5) \times 10^{13}$~h$^{-5/2}M_\odot$,  the
resulting gas-to-total fraction is
$M_{gas} / M_{tot} = (0.06 \pm 0.04)$~h$^{-3/2}$ if one adopts the
cluster mass estimated using the X-ray data  and
$M_{gas} / M_{tot} = (0.04 \pm 0.02)$~h$^{-3/2}$
if one uses the mass estimate derived from the lensing estimate.

Estimating the total stellar mass using a mean mass-to-light ratio of
$M/L = 8$~h, we find that it's contribution to the total baryonic mass is
negligible in comparison to the mass in the intracluster gas.  The above
values of $M_{gas}/M_{tot}$ suggests that the baryonic fraction in A2218,
within the central $\sim 400$~h$^{-1}$ is
$f_b \approx (0.04 \pm 0.02)$~h$^{-3/2}$.  Comparing this value to
the mean baryonic fraction of the universe expected from nucleosynthesis
arguments  (e.g.~Walker \etal~1991) of
$f_b \leq 0.015 \; (\Omega$h$^2)^{-1}$
suggests that the value of the cosmological density parameter is
$\Omega \simeq 0.3$.

Furthermore, adding the light in the field associated with {\it all} the
galaxies gives a lower bound on the mass-to-light ratio in A2218.  At the
fiducial radius,
we find that $(M/L_B) = (440 \pm 80)$~h$(M_\odot / L_{\odot B})$.
If we use only the red light associated with the bright galaxies in the
cluster  color sequence, we obtain
$(M/L_B) = 570$~h$(M_\odot / L_{\odot B})$. Here,
we have used the lensing mass estimate and have derived the $M/L$
ratio using the total projected mass and luminosity.  This
ratio remains roughly constant for $r=180^{\prime \prime}$ out to the edge
of our survey since both the projected light and the surface mass
density radial profiles
vary with radius in a similar fashion.  Converting the above $M/L$ into an
estimate of the cosmological density parameter once again yields
$\Omega \simeq 0.3$.

The mean surface mass density used to determine the weak lensing mass
estimate is, however, calculated by subtracting the mean in the reference
annulus extending from 3\farcm5 to
6\farcm2.  Only if the cluster mass distribution drops off sharply beyond
$\sim 400$~h$^{-1}$kpc, then the quoted value of the mass represents the
actual value of the projected mass internal to $\sim 400$~h$^{-1}$kpc.
If, however, the cluster mass
distribution extends beyond $\sim 400$~h$^{-1}$kpc (and hence,
into the reference annulus), then the above value is an underestimate.
There are indications that the latter is true; for example, the X-ray
surface brightness profile extends smoothly out to $\sim 1$~h$^{-1}$Mpc
and the projected cluster mass continues
to rise at least out to $\sim 600$~h$^{-1}kpc$.
If this is indeed so, then the actual value of the projected mass interior to
$\sim 400$~h$^{-1}$kpc could potentially be much higher.
We must caution, however, that this is a highly model-dependent scenario:
the isothermal model gives the actual mass being a factor $\sim 1.6$
higher than our quoted value, while a steeper, Hernquist-type profile gives
only a factor of $\simeq 1.2$.

It is tempting to speculate on the consequences of the cluster
following an $r^{-1}$ surface mass density profile in the control
annulus. In this scenario, we have the following
consequences:  First, the
lensing mass and the X-ray mass estimates would be discrepant by a
factor of $\sim 1.6$, with the lensing mass estimate being higher.
Interestingly, the level of discrepancy is
similar to that in the central regions of the cluster, as discussed by
Miralda-Escud\'e  \& Babul (1995) and Kneib \etal~(1995).  Second, the
gas-to-total mass ratio would drop to $f_b = (0.02 \pm 0.01)$h$^{-3/2}$,
a result that is consistent with
the nucleosynthesis value in an $\Omega=1$ universe.  Third, the
mass-to-light ratio would also increase to a value as large as $M/L \sim 900$,
implying a high value of $\Omega$.

We can, at this point, speculate at length on whether or not the X-ray and
lensing mass estimates are indeed discrepant away from the cluster center,
and on whether our analysis of A2218 supports the case for high or low
$\Omega$. All these issues, however, can be
settled observationally by directly probing the weak lensing distortions
out to even larger radii and establishing whether or not the mass distribution
associated with A2218 drops off sharply beyond a projected distance of
$500$~h$^{-1}$Mpc.  This exciting prospect
is now possible with the advent of large-format CCD arrays; for example,
the MOCAM device at CFHT has a 14$^\prime$ prime focus field.  As it comes
into its own, the weak lensing technique offers a unique opportunity to
resolve many of the outstanding puzzles regarding the dark matter content
and distribution in the universe.

\acknowledgments{It is a pleasure to acknowledge most valuable assistance
from Richard Griffiths and the MDS team. We are greatful to Megan
Donahue for enlightening discussions on the X-ray map of the cluster.
We thank Roser Pello for use of the compiled
redshifts measured for the cluster field. We greatfully acknowledge the
redshift data and extrapolation to faint magnitudes by Simon Lilly,
Caryl Gronwall, and David Koo.}

\end{document}